\documentclass[draftcls,onecolumn,12pt]{IEEEtran}
\pdfoutput=1
\ifCLASSINFOpdf
\else
\fi

\ifCLASSOPTIONcompsoc
\usepackage[caption=false,font=normalsize,labelfont=sf,textfont=sf]{subfig}
\else
\usepackage[caption=false,font=footnotesize]{subfig}
\fi

\usepackage{cite}
\usepackage{bm}
\usepackage{url}
\usepackage[dvipsnames]{xcolor}
\usepackage{amsmath}
\usepackage{mathtools}
\usepackage[]{amssymb}
\usepackage{amsthm}
\usepackage{float}
\usepackage{dsfont}
\usepackage{array}
\usepackage{booktabs}
\usepackage{multirow}
\usepackage{footmisc}
\usepackage{textcomp,gensymb}
\usepackage{graphicx}
\usepackage[para,flushleft]{threeparttable} 
\usepackage{multirow}	
\usepackage{dcolumn} 

\usepackage[ruled]{algorithm2e}
\usepackage{xifthen} 
\usepackage{enumitem}

\makeatletter
\newcommand{\removelatexerror}{\let\@latex@error\@gobble}
\makeatother

\DeclareMathOperator{\diag}{diag}

\newif \ifwebcolor
\webcolortrue

\def \ba {\begin{array}}
\def \ea {\end{array}}
\def \benu {\begin{enumerate}}
\def \eenu {\end{enumerate}}
\def \bdes {\begin{description}}
\def \edes {\end{description}}

\def \bitem {\begin{itemize}}
\def \eitem {\end{itemize}}

\def \bfl {\begin{flushleft}}
\def \efl {\end{flushleft}}
\def \bfr {\begin{flushright}}
\def \efr {\end{flushright}}

\def \beq {\begin{equation}}
\def \eeq {\end{equation}}
\def \bqa {\begin{eqnarray}}
\def \eqa {\end{eqnarray}}
\def \bqa* {\begin{eqnarray*}}
\def \eqa* {\end{eqnarray*}}

\def \bal {\begin{align}}
\def \eal {\end{align}}

\DeclareMathOperator*{\argmin}{arg\,min}

\begin{document}
%
\title{
A Message Passing Based Average Consensus 
Algorithm for Decentralized Frequency and Phase Synchronization in Distributed Phased Arrays
}

\author{\IEEEauthorblockN{Mohammed Rashid, ~\IEEEmembership{Member,~IEEE}, and 
Jeffrey A. Nanzer, ~\IEEEmembership{Senior Member,~IEEE}}}

\maketitle
\thispagestyle{empty}
\pagestyle{empty}

\def\bda{\mathbf{a}}
\def\bdd{\mathbf{d}}
\def\bde{\mathbf{e}}
\def\bdf{\mathbf{f}}
\def\bdg{\mathbf{g}} 
\def\bdh{\mathbf{h}}
\def\bdm{\mathbf{m}}
\def\bds{\mathbf{s}} 
\def\bdn{\mathbf{n}}
\def\bdu{\mathbf{u}}
\def\bdv{\mathbf{v}}
\def\bdw{\mathbf{w}} 
\def\bdx{\mathbf{x}} 
\def\bdy{\mathbf{y}} 
\def\bdz{\mathbf{z}}
\def\bdA{\mathbf{A}}
\def\bdC{\mathbf{C}}
\def\bdD{\mathbf{D}} 
\def\bdF{\mathbf{F}}
\def\bdG{\mathbf{G}} 
\def\bdH{\mathbf{H}}
\def\bdI{\mathbf{I}}
\def\bdJ{\mathbf{J}}
\def\bdU{\mathbf{U}}
\def\bdX{\mathbf{X}}
\def\bdK{\mathbf{K}}
\def\bdL{\mathbf{L}}
\def\bdQ{\mathbf{Q}_n}
\def\bdR{\mathbf{R}}
\def\bdS{\mathbf{S}}
\def\bdV{\mathbf{V}}
\def\bdW{\mathbf{W}}
\def\bdGamma{\bm{\Gamma}}
\def\bdgamma{\bm{\gamma}}
\def\bdalpha{\bm{\alpha}}
\def\bdmu{\bm{\mu}}
\def\bdSigma{\bm{\Sigma}_n}
\def\bdxi{\bm{\xi}}
\def\bdl{\bm{\ell}}
\def\bdLambda{\bm{\Lambda}}
\def\bdeta{\bm{\eta}}
\def\bdPhi{\bm{\Phi}}
\def\bdtheta{\bm{\theta}}
\def\bdTheta{\bm{\Theta}}
\def\bddelta{\bm{\delta}}

\def\btau{\bm{\tau}}
\def\deg{\circ}

\def\tq{\tilde{q}}
\def\tbdJ{\tilde \bdJ}
\def\l{\ell}
\def\bdzero{\mathbf{0}} 
\def\bdone{\mathbf{1}} 
\def\Exp{\mathbb{E}} 
\def\exp{\text{exp}} 
\def\ra{\rightarrow}

\def\R{\mathbb{R}} 
\def\C{\mathbb{C}} 
\def\CN{\mathcal{CN}} 
\def\N{\mathcal{N}} 
\begin{abstract}
We consider the problem of decentralized frequency and phase synchronization in distributed 
phased arrays via local broadcast of the node electrical states.
Frequency and phase synchronization between nodes in a distributed array is necessary to support beamforming, but 
due to the operational dynamics of the local 
oscillators of the nodes, 
the frequencies and phases of their output signals undergo the random drift and jitter 
in between the update intervals. Furthermore, frequency and phase estimation 
errors contribute to the total phase errors, leading to a residual phase error in the array that degrades coherent operation.
Recently, a classical decentralized frequency and phase synchronization algorithm based on consensus averaging was proposed with which 
the standard deviation of the residual phase errors upon convergence were reduced to $10^{-4}$ degrees for internode update intervals of $0.1$ ms, however this was obtained for arrays with at least $400$ nodes and a high connectivity ratio of $0.9$.  
In this paper, we propose a message passing based average 
consensus (MPAC) algorithm to improve the synchronization of the electrical states of the nodes in distributed arrays.
Simulation results show that 
the proposed MPAC algorithm significantly reduces the residual 
phase errors to about $10^{-11}$ degrees, 
requiring only $20$ moderately connected nodes in an array. Furthermore, MPAC converges faster 
than the DFPC-based algorithms particularly for the larger arrays with a moderate connectivity. 
\end{abstract}


\section{Introduction}

Distributed phased arrays (DPAs) are collections of separate antenna systems that are wirelessly coordinated to perform coherent operations such as beamforming. 
When compared to the large single-platform architecture that uses 
analog feed networks and a single transceiver 
chain to drive the antennas, this distributed architecture brings 
several advantages to wireless applications, including higher signal power at the destination, improved spatial diversity, improved adaptability to changing environments, 
higher resistance to the overall system failure, and ease in scalability of the system 
\cite{Nanzer_Survey_2021, OCDA_2017}.  
Each node in a DPA has its own transceiver chain with an independent local oscillator. 
When free running, the signals produced by each oscillator undergo random drift and jitter
over time that introduces a decoherence between the signals emitted by the array \cite{Serge_Access_2021, 
Phase_noise_2000}. 
Existing methods that can be used 
for node synchronization purposes may be classified as either suitable for a closed-loop system or 
an open-loop system. In a closed-loop system, the nodes use a feedback from the destination, 
e.g., the received signal strength, to tune their oscillators 
until a significant coherence level is achieved at the destination \cite{1_bit_feedback, 3_bit_feedback, retrodirective_2016}. The benefit of this approach is that little coordination is explicitly required between nodes; however, closed-loop systems cannot arbitrarily beamform, thus operations like radar and sensing are not feasible.
On the other hand, in an open-loop system, the nodes do not use any feedback from the destination and 
synchronize their oscillators by exchanging signals with each other. 
Therefore, open-loop methods as proposed 
in \cite{OCDA_2017, Sean_2020, Sean_TMTT_2019} can also be used for the radar applications \cite{Schmid_radar}, however they require more synchronization than closed-loop systems.

In \cite{rashid2022frequency}, we proposed a decentralized frequency and phase consensus (DFPC) algorithm for open-loop DPAs
in which the nodes share their frequencies and phases with each other through a local 
broadcast of their signals, and iteratively update these parameters by computing a weighted average of 
the received values. Simulation results in \cite{rashid2022frequency} 
show that with the DFPC-based algorithms, 
the standard deviation of the residual phase errors upon convergence 
can be reduced to about $10^{-4}$ degrees, for a practical update interval of $0.1$ ms, that requires 
at least $400$ nodes in an array having a high connectivity ratio of $0.9$ 
(see Fig. $9$ in \cite{rashid2022frequency}). Essentially, the DFPC algorithm is based on 
the average consensus algorithm of \cite{Fast_MC_2004} and thus begins 
with constructing a Markov chain (MC) with a doubly-stochastic transition matrix (a.k.a  
the mixing matrix or the weighting matrix).  
The DFPC algorithm is started with an arbitrary distribution over the frequencies 
and phases, and it progresses by mixing the frequencies and phases in each iteration 
until convergence where the synchronization is also achieved.
However, with the MC-based mixing matrix, the synchronization 
is achieved only asymptotically, and therefore, DFPC takes a large number of 
iterations to converge even for larger arrays with a moderate 
connectivity. To improve its convergence speed, a better mixing matrix can be constructed with a 
smaller second largest eigenvalue; however, this requires global connectivity 
information at each node, which is not generally available in dynamic distributed arrays.
In general, a large number of convergence iterations 
introduces a delay in achieving the synchronized state that 
is intolerable particularly when low-powered nodes are considered. Furthermore, an improved 
synchronization level between the nodes is also highly desirable to ensure a high gain coherent operation 
at the destination \cite{OCDA_2017}.

Recently, a Gaussian belief propagation \cite{BP_algo} 
based on an average consensus algorithm was proposed in \cite{BP_like1, BP_like2} 
with the motivation that computing the marginals 
can be equated to solving an average consensus problem \cite{Zhang_2020}.  
Thus, this algorithm is based on an alternative approach where the messages containing the 
coarse weighted averages and the weights-sum information are propagated through 
the network to solve the consensus problem. 
In this paper, we extend the 
algorithm in \cite{BP_like1, BP_like2} to solve the frequency and phase synchronization problem 
in a distributed phased array. To this end, 
we take into account the 
frequency drifts and phase jitters induced by the oscillators, as well as the frequency and phase estimation 
errors at the nodes due to the local broadcasting of the signals. A message passing based average consensus 
(MPAC) algorithm is developed in which the nodes iteratively exchange messages with their neighboring nodes to 
reach the consensus, i.e., synchronization in frequency and phase. Unlike the previously proposed DFPC 
and Kalman filtering based DFPC (KF-DFPC) algorithms in 
\cite{rashid2022frequency}, the MPAC 
algorithm does not require the network connectivity information to assign weights to the nodes. 
Simulation results show that compared to DFPC and KF-DFPC, MPAC significantly reduces the residual 
phase errors upon convergence to about $10^{-11}$ degrees 
with only $20$ moderately connected nodes in the array and irrespective of the signal to noise ratio (SNR) 
of the signals. 
Furthermore, MPAC takes fewer iterations for convergence than the DFPC-based algorithms particularly 
for larger arrays with a moderate connectivity. 

The rest of this article is outlined as follows. 
Section \ref{Freq_Phase_DA} formulates the decentralized frequency and phase 
synchronization problem in a DPA, proposes an MPAC algorithm to synchronize these 
parameters across the array, and 
theoretically analyzes the residual phase errors. 
Simulation results are included in Section \ref{MPAC_sims} wherein the synchronization 
performance of MPAC is investigated and compared to the DFPC algorithm. 
Finally, Section \ref{conclusion_section} concludes this work.

%

\section{Decentralized Frequency and Phase Synchronization in Distributed 
Phased Arrays}\label{Freq_Phase_DA}

Consider a group of $N$ nodes that are connected together 
with bidirectional communication links to form a distributed phased array. 
The network of these $N$ nodes can be represented by an undirected graph $\mathcal{G}=(\mathcal{V},\mathcal{E})$
in which $\mathcal{V}=\{1,2,\ldots,N\}$ represents the set of vertices, 
and $\mathcal{E}=\{(m,n)\colon m,n\in \mathcal{V}\}$ denotes the set of all undirected edges in the graph. 
We assume that the nodes are communicating with each other via a local broadcast of signals 
to synchronize their frequencies and phases across the array. 
Let the 
signal generated by the $n$-th node in iteration $k$ over the time duration $T$ be given by 
$s_n(t)=e^{j\left(2\pi f_n(k) t+\theta_n(k)\right)}$, in which $f_n(k)$ and 
$\theta_n(k)$ represent the frequency and phase of the signal 
in the $k$-th iteration. In practice, these parameters in each iteration are 
influenced by the frequency drifts and phase jitters of the oscillators, and thus we model them 
in the $k$-th iteration as 
\begin{align}\label{freq_phase_process}
f_n(k)&=f_n(k-1)+\delta f_n\nonumber\\
\theta_n(k)&=\theta_n(k-1)+\delta \theta^f_n +\delta\theta_n,
\end{align}
in which $\delta f_n$ denotes the frequency drift of the oscillator at the update time, and 
parameter $\delta \theta^f_n$ represents the phase due to its 
temporal variation over the time duration $T$ which is given by 
$\delta\theta^f_n=-\pi T\delta f_n$ \cite{Serge_Access_2021, rashid2022frequency}, and
$\delta \theta_n$ models the phase jitter of the oscillator 
at the $n$-th node~\cite{Serge_Access_2021}. 
The frequency and phase evolutions in \eqref{freq_phase_process} start with the initial values 
$f_n(0)$ and $\theta_n(0)$, respectively. We assume that the initial frequency of the 
$n$-th node is normally distributed as 
$f_n(0)\sim\N(f_c,\sigma^2)$ where $f_c$ 
is the carrier frequency, $\sigma=10^{-4}f_c$ denotes a crystal clock accuracy of $100$ parts 
per million (ppm), and the initial phase is 
uniformly distributed as $\theta_n(0)\sim \mathcal{U}(0,2\pi)$. As the nodes in the array 
share their frequencies and phases with each other 
through a local broadcast of their signals,  
the shared values are also influenced by the estimation errors. Thus, the frequency and phase of the $n$-th 
node observed in the iteration $k$ are written as 
\begin{align}\label{freq_phase_process_1}
 \hat{f}_n(k)&=f_n(k)+\varepsilon_f\nonumber\\
\hat{\theta}_n(k)&=\theta_n(k)+\varepsilon_\theta,
\end{align}
where $\varepsilon_f$ and $\varepsilon_\theta$ represent the frequency and phase estimation errors 
at the node. 

Next we describe the statistical modeling of the frequency drift and phase jitter of the 
oscillators, as well as the frequency and phase estimation errors at the nodes as follows. 
To begin, the frequency drift of the oscillator at the $n$-th node is modeled as 
$\delta f_n\sim \N\left(0,\sigma^2_f\right)$ in which the standard deviation of 
the frequency drift $\sigma_f$ can 
be set equal to the Allan deviation (ADEV) of the oscillator~\cite{Serge_Access_2021}. 
The ADEV is defined as the standard deviation of the averaged fractional frequency errors computed 
over multiple shifted time intervals. 
We model the ADEV as $\sigma_f=f_c\sqrt{\frac{\beta_1}{T}+\beta_2 T}$ with  
$\beta_1$ and $\beta_2$ depend on the design of the oscillator that we define as 
$\beta_1=\beta_2=5\times 10^{-19}$ \cite{rashid2022frequency}. 
The phase jitter 
of the 
oscillator at the $n$-th node is modeled as $\delta\theta_n\sim \N(0,\sigma_\theta)$ 
where its standard deviation is defined as 
$\sigma_\theta=\sqrt{2\times 10^{A/10}}$ in which $A$ represents the integrated phase noise power of an oscillator. The parameter $A$ is defined as the $\log_{10}$ of the total area under the entire curve 
of the phase noise profile of an oscillator. 
Herein, we set $A=-53.46$ dB that models a typical high phase noise voltage controlled oscillator  
\cite{Serge_Access_2021, rashid2022frequency}. 
Finally, the frequency and phase estimation errors, i.e., 
$\varepsilon_f$ and 
$\varepsilon_\theta$, are also modeled as normally distributed with zero mean and standard 
deviations $\sigma^m_f$ and $\sigma^m_\theta$, respectively. As the focus here is on 
the synchronization problem, we set these standard deviations equal to the Cramer-Rao 
lower bounds (CRLBs)
that are derived in \cite{Richards_radar}. Thus we set
$\sigma^m_f=\sqrt{\frac{6}{(2\pi)^2 L^3 \text{SNR}}}$
and 
$\sigma^m_\theta= \frac{2L^{-1}}{ \text{SNR}}$ in which 
the SNR denotes 
the signal to noise ratio of the received signals, 
and $L=Tf_s$ represents the number of samples 
collected over the observation window of length $T$ with sampling frequency $f_s$. 
Note that these CRLBs 
can be achieved by an unbiased and efficient estimators, for e.g., the FFT-based maximum likelihood 
estimators described in \cite{Liao2011}, assuming a large number of samples are available for the estimation purposes. 

For an ideal synchronization of nodes, the total phase error defined as 
$\delta \phi_n=2\pi \delta f_n T+ 2\pi\varepsilon_f T
+\delta \theta^f_n+\delta\theta_n+\varepsilon_\theta$ 
must be zero for all the nodes across the array. 
However, in practice the residual error will not converge to zero due to propagation delays of the signals and the continual drift of the oscillators; thus 
we define synchronization of the frequency and phase of the nodes in the array 
when the standard deviation of the total phase errors $\delta \phi_n$ 
satisfies:
\begin{align}
\sigma_\phi&=\sqrt{\frac{1}{N-1}\sum^N_{n=1} \mid \delta \phi_n-\bar{\phi}\mid^2}\leq \eta,
\end{align}
in which $\eta$ represents some pre-defined threshold, and $\bar{\phi}$ denotes the average value of the total 
phase errors. It is established in Fig. 4 in 
\cite{OCDA_2017}, that at least $90\%$ of the ideal 
coherent gain can be achieved at the destination if 
the standard deviation $\sigma_\phi$ is below the threshold $\eta=18^\deg$. In other words, any $\eta$ below 
$18^\deg$ guarantees high coherent gain at the destination. 

\subsection{Message Passing Based Average Consensus Algorithm}

We assume that each node iteratively exchanges its frequency and phase only with 
its neighboring nodes and updates these parameters in each iteration by computing a 
weighted average of the shared values. 
This local sharing of the information between the nodes 
enables the use of a fully decentralized (distributed) algorithm that is easily scalable 
as the required resources per node for its implementation -- 
for instance, the memory storage, the computational power, and the bandwidth -- 
are mainly controlled by the average number of neighbors per node in a network. 

\begin{figure}[tp]
	\centering
		\includegraphics[width=0.48\textwidth]{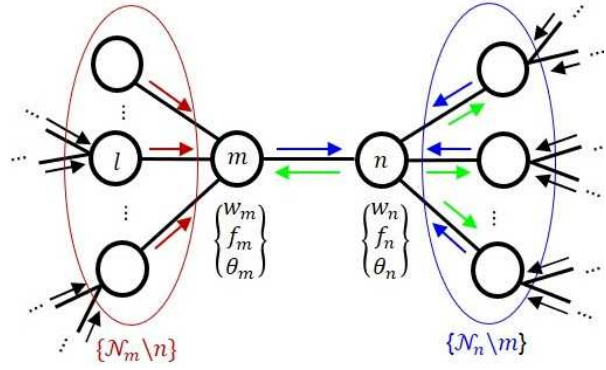}
	\caption{A portion of an undirected graph depicting the flow of 
	messages needed to update the frequency and phase of node $n$ in the $k$-th iteration. 
	The red-colored arrows between nodes $l$ and $m$ (for each $l\in\{\N_m\backslash n\}$) represent the messages 
	$\left(\mu^f_{l\ra m}(k-2),\mu^\theta_{l\ra m}(k-2), s_{l\ra m}(k-2)\right)$ computed in $(k-2)$-th iteration. 
	These messages are used by node $m$ to send out the messages 
	$\left(\mu^f_{m\ra n}(k-1),\mu^\theta_{m\ra n}(k-1), s_{m\ra n}(k-1)\right)$ to node $n$ 
	in the $(k-1)$-st iteration (following 
	\eqref{upd_mesg_f} and \eqref{upd_mesg_theta}) 
	which is shown with the blue-colored arrows. Similarly, the 
	nodes in set $\{\N_n\backslash m\}$ also follow the same procedure to compute these blue-colored 
	messages in $(k-1)$-st iteration using the messages from the previous iteration. 
	Finally, node $n$ uses \eqref{f_upd} and 
	\eqref{theta_upd} to update its frequency and phase by combining all 
	these blue-colored messages from its neighboring nodes in the set $\N_n$. It then sends out the new messages 
	$\left(\mu^f_{n\ra m}(k),\mu^\theta_{n\ra m}(k), s_{n\ra m}(k)\right)$ to all $m\in \N_n$ 
	as shown by the green-colored arrows.}
		\label{fig:msgs_flow}
\end{figure}

Each node $n$ in the array has a weight $w_n$ assigned to it, and 
let $f_n(k)$ and $\theta_n(k)$ represent its updated frequency and phase in the $k$-th iteration. 
As shown in Fig. \ref{fig:msgs_flow}, 
we assume that, for each $m\in \N_n$, $\mu^f_{m\ra n}(k-1)$ and $\mu^\theta_{m\ra n}(k-1)$ denote the 
messages sent from node $m$ to node $n$ in the $(k-1)$-st iteration, that represents the
coarse frequency and phase weighted averages, respectively, 
computed at node $m$ by using all the values of its neighboring nodes from the previous iteration 
except the shared values from node $n$. 
Likewise, let $s_{m\ra n}(k-1)$ denote 
the sum of the weights of the neighboring nodes of node $m$, computed in the $(k-1)$-st iteration, 
excluding the weight $w_n$ of node $n$. 
Node $m$ uses $s_{m\ra n}(k-1)$ to compute 
the coarse weighted averages $\mu^f_{m\ra n}(k-1)$ and $\mu^\theta_{m\ra n}(k-1)$ in iteration $k-1$ and thus passes that 
scalar to node $n$ as well. 
Node $n$ receives these messages from all its neighboring nodes, then it 
updates its frequency and phase values in the $k$-th iteration by 
combining all the received values as follows
\begin{align}\label{f_upd}
f_n(k)&=\frac{w_n \hat{f}_n(k)+\sum_{m\in \N_n} s_{m\ra n}(k-1)\mu^f_{m\ra n}(k-1)}
{s_n(k)},\\ 
\theta_n(k)&=\frac{w_n \hat{\theta}_n(k)+\sum_{m\in \N_n} s_{m\ra n}(k-1)\mu^\theta_{m\ra n}(k-1)}
{s_n(k)},\label{theta_upd}
\end{align}
in which $s_n(k)=w_n+\sum_{m\in \N_n} s_{m\ra n}(k-1)$ and $\N_n$ is the set of neighboring nodes of node $n$.
		%

Next node $n$ sends out the updated messages (the 
coarse frequency and 
phase weighted averages) to each of its $m$ neighbors for all $m\in\N_n$. These messages are computed as 
\begin{align}\label{upd_mesg_f}
\mu^f_{n\ra m}(k)
&=\frac{w_n \hat{f}_n(k)+\sum_{m\in \{\N_n\backslash{m}\}} s_{m\ra n}(k-1)\mu^f_{m\ra n}(k-1)}
{w_n+\sum_{m\in\{\N_n\backslash{m}\}} s_{m\ra n}(k-1)},
\end{align}
and
\begin{align}\label{upd_mesg_theta}
\mu^\theta_{n\ra m}(k)
&=\frac{w_n \hat{\theta}_n(k)+\sum_{m\in \{\N_n\backslash{m}\}} s_{m\ra n}(k-1)\mu^\theta_{m\ra n}(k-1)}
{w_n+\sum_{m\in\{\N_n\backslash{m}\}} s_{m\ra n}(k-1)},
\end{align}
where $\{\N_n\backslash{m}\}$ is the set of neighboring nodes of node $n$ 
excluding node $m$. Note that the message $s_{n\ra m}(k)$ is defined as 
$s_{n\ra m}(k)=f_\gamma\left(w_n+\sum_{m\in\{\N_n\backslash{m}\}} s_{m\ra n}(k-1)\right)$ where the  
function $f_\gamma(x)=\frac{\gamma x}{\gamma+x}$ with $\gamma\gg 1$ 
is used to ensure that the proposed MPAC algorithm converges 
in case of both acyclic as well as cyclic networks \cite{BP_like1}. 
The updated coarse weighted averages from \eqref{upd_mesg_f} and \eqref{upd_mesg_theta} are then used 
in the next iteration by the $m$-th node for updating its 
frequency and phase values following \eqref{f_upd} and \eqref{theta_upd}. 
The above steps are repeated at each node in the graph 
in every iteration until the convergence is achieved. 
This message passing based average consensus (MPAC) algorithm is described in detail in Algorithm 1. 

\begin{algorithm}\label{algo_1}
  \footnotesize
\DontPrintSemicolon
\SetKwInput{KwPara}{Input}
\KwPara{$k=0$, define $w_n$. 
Next for each 
node $n$, and for each $m\in \N_n$, set 
$\mu^f_{m\ra n}(0)=f_c$, 
$\mu^\theta_{m\ra n}(0)=\pi$, 
and $s_{m\ra n}(0)=f_\gamma(w_m)$.}
\While{convergence criterion is not met} 
{
	$k=k+1$\\
For each node $n$:
\begin{enumerate}
\item[a)] Update the frequency $f_n(k)$ 
and phase $\theta_n(k)$ using \eqref{f_upd}\\ and 
\eqref{theta_upd}, respectively.
\end{enumerate}
then for each node $n$ and its each neighbor $m\in\N_n$:
\begin{enumerate}
\item[b)] Update the scale $s_{n\ra m}(k)$ using:\\
 $s_{n\ra m}(k)=f_\gamma\left(w_n+\sum_{m\in\{\N_n\backslash{m}\}} s_{m\ra n}(k-1)\right)$
\item[c)] Compute the new messages 
$\mu^f_{n\ra m}(k)$ and $\mu^\theta_{n\ra m}(k)$\\ using 
\eqref{upd_mesg_f} and \eqref{upd_mesg_theta}, respectively, 
and send them out to\\ node $m$. 
\end{enumerate}

}
\KwOut{$f_n(k)$ and $\theta_n(k)$ for all $n=1,2,\ldots,N$}
\caption{MPAC Algorithm}
\end{algorithm}

\subsection{Residual Phase Error Analysis} \label{resd_phase_analysis}
In this subsection, we theoretically examine the residual phase error of the proposed MPAC algorithm 
in the presence of the frequency and phase offset errors introduced at the nodes. To begin, 
the MPAC algorithm tends to solve the following optimization problem \cite{BP_like1}.
\begin{align}\label{opt_MPAC}
\argmin_\bdx \sum^N_{n=1} w_n|x_n-z_n(I)|^2+ \gamma\sum_{(m,n)\in\mathcal{E}} |x_m-x_n|^2,
\end{align}
in which $\bdx=[x_1, x_2,\ldots, x_N]^T$ is a consensus vector, 
$z_n(I)\in\{\hat{f}_n(I),\hat{\theta}_n(I)\}$ at iteration $I$ and thus correspondingly 
$x_n$ either represents a frequency consensus or phase consensus. 
$\gamma>0$ is the penalty parameter added to enforce the convergence where all $x_n$s become similar 
for a connected graph $\mathcal{G}$. 
Following Lemma 2 in \cite{BP_like1}, it can be easily shown that 
the objective 
function in \eqref{opt_MPAC} is strictly convex and its global minimum 
is given by $\bdx^\ast=(\gamma \bdL+\bdW)^{-1}\bdW\bdz(I)$ 
in which $\bdL$ is the $N\times N$ Laplacian matrix of $\mathcal{G}$, $\bdW$ is a diagonal matrix 
defined as $\bdW=\diag\{w_1,w_2,\ldots,w_N\}$, and $\bdz(I)=[z_1(I),z_2(I),\ldots,z_N(I)]^T$. Note that using 
\eqref{freq_phase_process} and \eqref{freq_phase_process_1}, we define $\bdz(I)=\bdz(I-1)+\bde_I$ where 
$\bde_I=[e_1(I),e_2(I),\ldots,e_N(I)]^T$ is offset error vector with $e_n(I)\sim\N(0,\sigma^2_e)$ in which 
$\sigma^2_e=\sigma^2_f+(\sigma^m_f)^2$ when $z_n(I)=\hat{f}_n(I)$, and 
$\sigma^2_e=(\pi T\sigma_f)^2+(\sigma^m_\theta)^2+\sigma^2_\theta$ when $z_n(I)=\hat{\theta}_n(I)$.
Using the backward recursion, the  
consensus vector $\bdx^\ast$ can be written as 
$\bdx^\ast=(\gamma \bdL+\bdW)^{-1}\bdW\bdz(0)+\sum^{I-1}_{i=0}(\gamma \bdL+\bdW)^{-1}\bdW\bde_{I-i}(0)$. 
The first term in this solution 
gives an average of the initial vector values and likewise the second terms gives the accumulated averaged errors. 
Essentially, the accumulated averaged errors becomes 
negligible for the large connected networks when larger $\gamma$ is used. 
Specifically, when $\gamma\ra \infty$ then \eqref{opt_MPAC} 
reduces to $\argmin_{x_n}\sum^N_{n=1}w_n|x_n-z_n(I)|^2$ as all the $x_n$ become similar. 
Then by taking the derivative of this new objective function with respect to $x_n$ and 
setting it to zero, we get $x^\ast_n=\frac{\sum^N_{n=1}w_n z_n(I)}{\sum^N_{n=1}w_n}$. 
In terms of the offset errors, $x^\ast_n$ 
can be written as 
$x^\ast_n=\frac{\sum^N_{n=1}w_n z_n(0)}{\sum^N_{n=1}w_n}+\frac{\sum^N_{n=1}w_n \sum^{I-1}_{i=0}e_n(I-i)}{\sum^N_{n=1}w_n}$. 
When $w_n=1$, the two summands compute the statistical means and because 
the sum of the offset errors $e_n(.)$
is normally distributed with zero mean, the errors are averaged out for larger networks 
when $\gamma\ra\infty$. This results in an improved synchronization performance of the MPAC algorithm. 

\section{Simulation Results}\label{MPAC_sims}
In this section, we investigate the frequency and phase synchronization 
performance of the proposed MPAC algorithm through 
simulations. To this end, we consider a network of $N$ nodes randomly generated 
with a connectivity $c$ which is defined as the ratio of the number of 
active edges in the network to the number of all possible edges ($N(N-1)/2$). 
Thus $c\in[0,1]$ and 
a higher value of $c$ implies a densely connected network, whereas a smaller value of $c$ 
means a sparsely connected network. Furthermore, the average number of 
connections per node are given by $D=c(N-1)$. We assume that the nodes transmit at a carrier 
frequency of $f_c=1 \text{ GHz}$ and use the sampling frequency of $f_s=10 \text{ MHz}$ 
to sample the received signals over $T=0.1$ ms interval. The weight of the $n$-th node in MPAC 
is set as $w_n=1$ and $\gamma=10^{12}$ to ensure improved synchronization between the nodes. To generate 
the figures in this section, the array network was randomly generated in each trial and the 
results were averaged over $10^3$ independent trials.

\begin{figure}[tp]
    \begin{minipage}{0.50\textwidth}
        \centering
\includegraphics[width=1.09\textwidth]{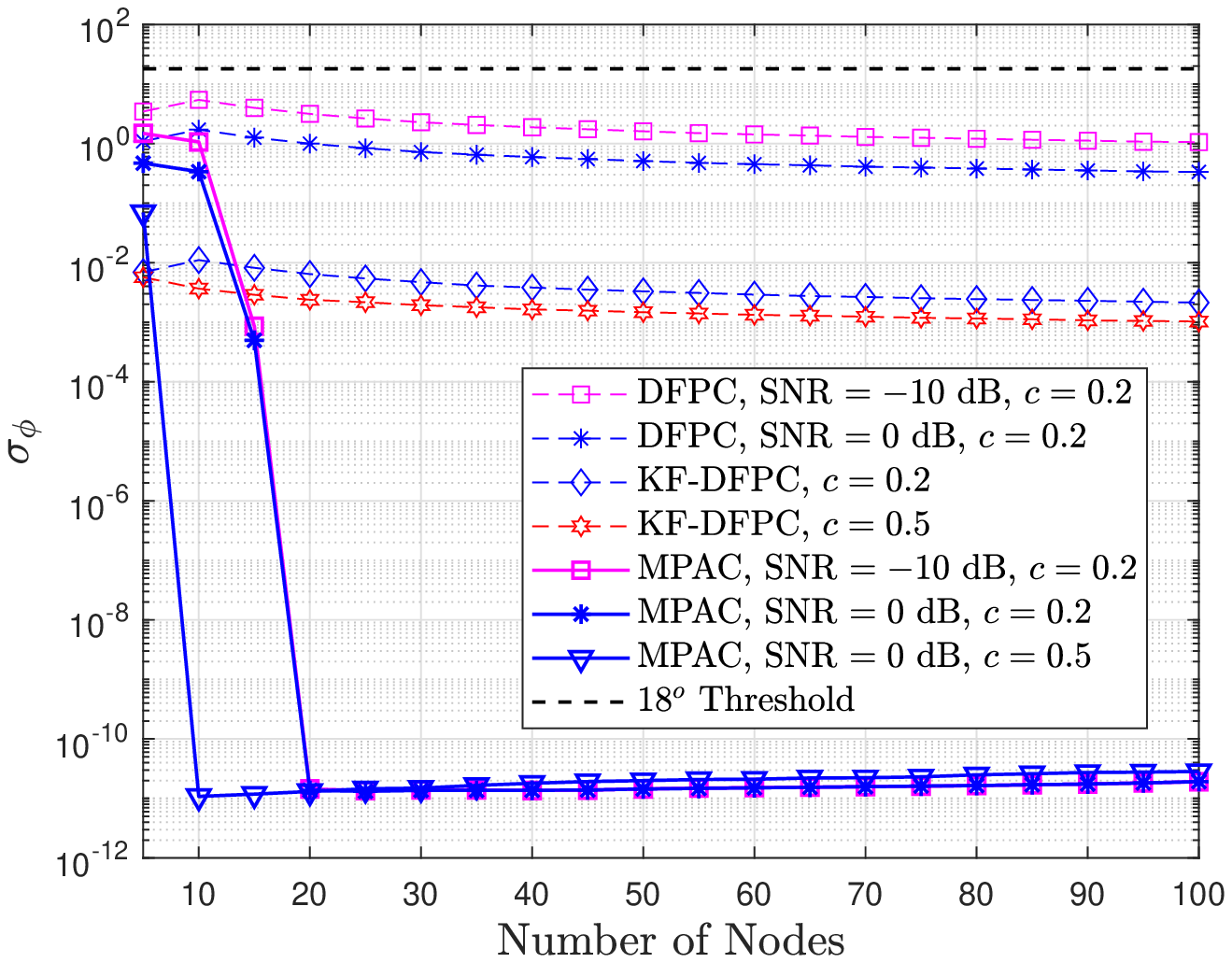}
	\caption{Standard deviation of the total phase errors of the MPAC and DFPC algorithms vs. 
	the number of nodes $N$ in the array when different SNR values and connectivity $c$ are considered.}
	\label{fig:sterr_vs_MPAC_OE}
		\end{minipage}\hspace{.01\linewidth}
    \begin{minipage}{0.50\textwidth}
        \centering
\includegraphics[width=1.09\textwidth]{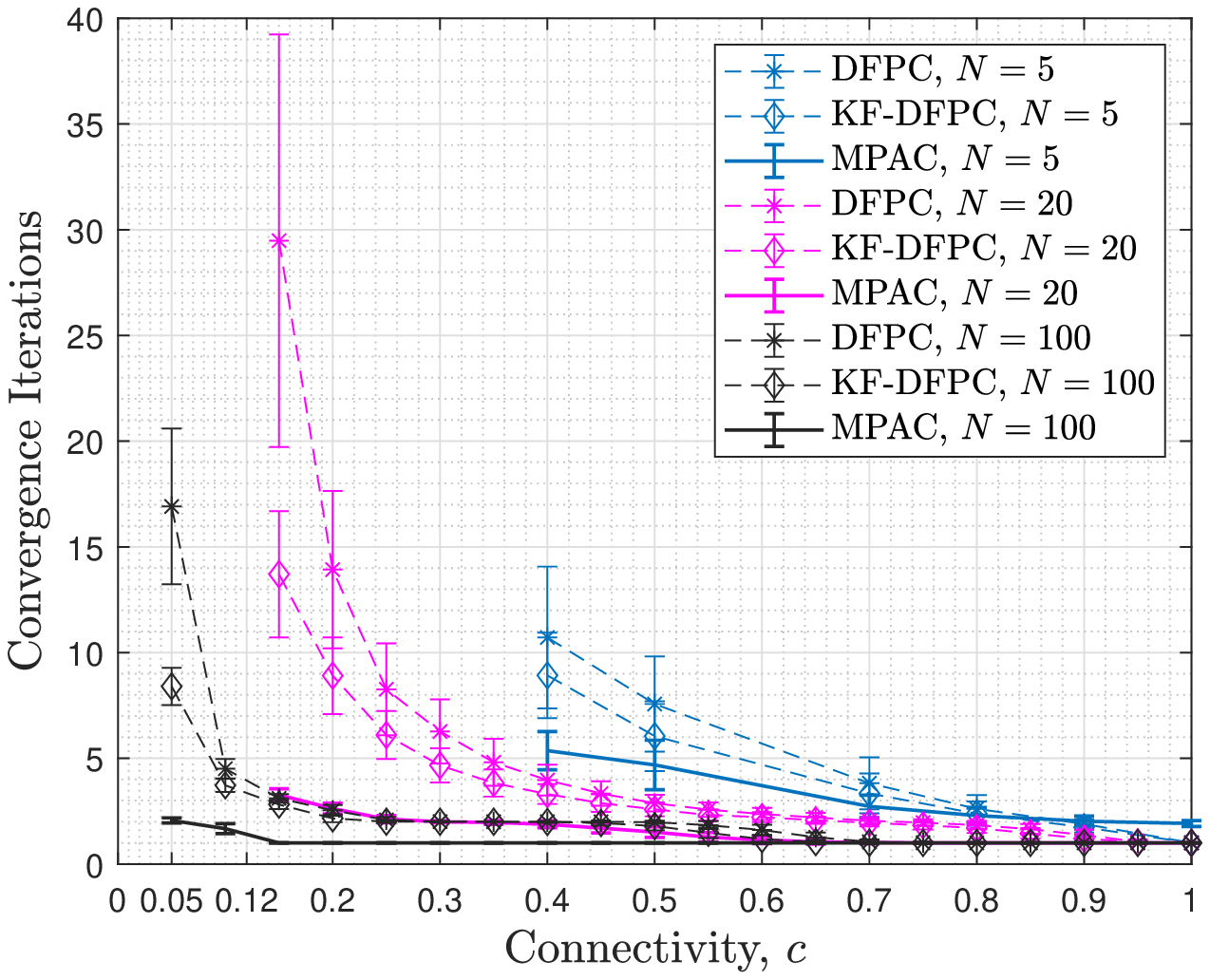}
	\caption{Convergence iterations of the MPAC and DFPC algorithms vs. 
	connectivity $c$ in the array when different number of nodes $N$ are considered for SNR $=0$ dB.}
	\label{fig:convg_vs_c_MPAC_OE}
		\end{minipage}
\end{figure}

In Fig. \ref{fig:sterr_vs_MPAC_OE}, we compare the standard deviation of the total 
phase errors $\delta\phi_n$ 
of the MPAC algorithm upon convergence to that of 
the DFPC and KF-DFPC algorithms proposed in \cite{rashid2022frequency} 
by varying the number of nodes in the array, the connectivity $c$ between the nodes, and the SNR 
of the received signals. 
Note that the minimum possible connectivity for $N=5$ nodes is $c=0.4$, whereas for $N\geq 10$ nodes we can set 
either $c=0.2$ or $0.5$. Furthermore, the performance of $\text{KF-DFPC}$ is independent of the SNR values as shown 
in \cite{rashid2022frequency}, and thus we illustrate here 
its performances for $c=0.2$ and $0.5$ values for the comparison purposes. 
This figure shows that with the increase in 
the number of nodes $N$ in the array, 
the standard deviation of the total phase errors decreases for all the algorithms. This is 
due to the increase in $D$ 
which assists in computing more accurate local averages at the nodes. However, 
the decrease is more rapid and significant for the MPAC algorithm as compared to the other algorithms 
for the larger $N$ values. Moreover, it is observed that while the performance of the 
DFPC algorithm improves with the increase in SNR due to the decrease in 
the estimation errors, on the other hand, the MPAC's performance 
is consistent at higher $N$ values irrespective of the SNR values. The KF-DFPC algorithm, which uses a Kalman filter, reduces the total phase 
errors for larger $N$ values but not as much as the MPAC algorithm. 
This improvement in the performance of MPAC is due to its 
averaging out of the errors as explained in Section \ref{resd_phase_analysis} where the averages are 
more unbiased for the larger $D$ values. 
In contrast, the residual phase errors of DFPC and KF-DFPC depend on the modulus 
of the second largest eigenvalue of the 
weighting matrix that is controlled by the $c$ value (as derived in 
Section III-A in \cite{rashid2022frequency}). Thus, the KF-DFPC's residual phase errors 
decreases with the increase in $c$, but not as much as the MPAC algorithm. 
 
Finally, in Fig. \ref{fig:convg_vs_c_MPAC_OE} we compare the convergence speeds of the 
three algorithms for different number of nodes in the array by varying the 
connectivity $c$ between the nodes for 
$\text{SNR}=0$ dB. This figure shows the average value and standard deviation of the $10^3$ 
samples using the errorbar plot. 
The threshold $\eta$ for convergence was set to $1^\deg$ which ensures high coherent gain operation 
at the destination \cite{OCDA_2017}. As expected, it is observed that the convergence rate of 
both algorithms improves with the increase in the connectivity $c$ between the nodes or 
the number of nodes $N$ in the array. However, the proposed MPAC algorithm takes 
considerably smaller number of convergence iterations for the moderately connected arrays 
with $c$ in $[0.05, 0.6]$. For e.g., for $N=20$ and $c=0.2$, DFPC takes $14$ iterations, KF-DFPC takes $9$ iterations, 
and MPAC takes $3$ iterations, whereas for $N=100$ and $c=0.05$, DFPC takes $17$ iterations, 
KF-DFPC takes $8$ iterations, and MPAC takes $2$ iterations.  

\section{Conclusions}\label{conclusion_section}
The frequency and phase synchronization problem is an essential bottleneck for 
leveraging the benefits of the distributed phased array, particularly when 
the frequency and phase offset errors are introduced at the nodes. 
We developed 
a decentralized MPAC algorithm that synchronizes these parameters across the array 
through a local propagation of messages between the nodes. 
Simulation results show that our proposed 
MPAC algorithm significantly 
reduces the residual phase errors upon convergence as compared to the DFPC-based algorithms. 
In particular, MPAC reduces the standard deviation of the 
total phase errors to about $10^{-11}$ degrees with only $20$ moderately connected nodes in the array and 
irrespective of the SNR of the received signals. Moreover, it converges in a fewer 
iterations as compared to the DFPC-based algorithms. 

\bibliographystyle{IEEEtran}
\bibliography{References}
\end{document}